\documentclass[aps,prd,floatfix,showpacs,10pt,nofootinbib]{revtex4-1}

\usepackage{amsmath,amssymb,amsfonts,amsthm} 
\usepackage{color}
\usepackage{float}
\usepackage{graphicx}
\usepackage{pstricks,pstricks-add}
\usepackage{pst-plot}
\usepackage[scriptsize,nooneline,hang]{caption}
\usepackage[hang,nooneline,scriptsize]{subfigure}

\definecolor{dark-green}{rgb}{0,0.7,0}
\definecolor{dark-blue}{rgb}{0,0.2,0.5}
\definecolor{med-blue}{rgb}{0,0.7,1}
\definecolor{mblue}{rgb}{0,0.2,1}
\definecolor{cnc}{rgb}{0.8,0,0}
\definecolor{light-red}{rgb}{1,0.8,0.8}
\definecolor{dark-yellow}{rgb}{1,0.8,0}
\definecolor{light-blue}{rgb}{0.8,0.9,1}
\definecolor{grey}{rgb}{0.211,0.211,0.211}
\definecolor{verylight-blue}{rgb}{0.93,0.95,1}
\definecolor{light-yellow}{rgb}{1,0.9,0.8}

\begin{document}

\title{Critical phenomena of charged Einstein-Gauss-Bonnet
  black holes \\ with charged scalar hair}

\author{Yves Brihaye $^{(a)}$}

\author{Betti Hartmann $^{(b)}$ }

\affiliation{
$(a)$ Physique-Math\'ematique, Universit\'e de Mons-Hainaut, 7000 Mons, Belgium\\
$(b)$ Instituto de F\'isica de S\~ao Carlos (IFSC), Universidade de S\~ao Paulo (USP), CP 369,
13560-970 , S\~ao Carlos, SP, Brazil}
\date{\today}

\begin{abstract}
Einstein-Gauss-Bonnet-gravity (EGB) coupled minimally to a $U(1)$ gauged, massive scalar field possesses
-- for appropriate choices of the $U(1)$ charge -- black hole solutions that carry charged scalar hair if 
the frequency of the harmonic time-dependence of the scalar field is equal to the upper bound on the superradiant frequency. 
The existence of these solutions has first been discussed in \cite{Grandi:2017zgz}. 
In this paper, we demonstrate that the critical value of the scalar charge results from the requirement
of non-extremality of the charged black hole solutions and the fact that the scalar field should not escape to 
infinity. Moreover, we investigate the hairy black holes in more detail and demonstrate that the branch of 
these solutions joins the branch of the corresponding charged EGB black hole for vanishing scalar field, but is {\it not} connected to the branch of boson stars in the limit of vanishing horizon radius. This indicates that
it is unlikely that these black holes 
appear from the collapse of the corresponding boson stars. Finally, we prove a No-hair theorem for charged scalar fields
with harmonic time-dependence for static, spherically symmetric, asymptotically flat electro-vacuum black holes in $d$ space-time dimensions and hence
demonstrate that the GB term is crucial for the existence of the hairy black holes discussed in this paper.

\end{abstract}

\maketitle

\section{Introduction} 
The first direct observation of gravitational waves (GW) by the LIGO detector in 2015 \cite{ligo1} has not only allowed the confirmation of a prediction
of General Relativity (GR) made by Einstein in 1918 \cite{einsteinGW}, but also hints to the actual astrophysical existence of 
objects that exist -- in contrast to gravitational waves -- in the non-linear regime of the theory: black holes (see e.g. \cite{chandrasekhar}).
All gravitational wave signals that do not match corresponding observations in the electromagnetic spectrum (see e.g. \cite{ligo2} for
a merger of two neutron stars and the corresponding observation in gamma rays \cite{GWgamma1}) are in perfect agreement with templates generated from simulations of the merger of two rotating, vacuum black holes \cite{ligo3} within the framework of GR \cite{ligo4}. In particular, the ringdown phase of the final -- initially perturbed -- rotating, vacuum black hole
is interesting from the point of view of black hole properties. The ringdown is dominated by the quasi-normal modes (QNM) of the black hole \cite{QNM}. The complex valued frequency of each of these modes is solely given in terms of the mass $M$ and the angular momentum $J$ of the rotating, vacuum black hole  (see e.g. \cite{Berti:2009kk}
and references therein).  This leads automatically to one of the important conjectures, the so-called {\it No hair conjecture} \cite{Ruffini:1971bza} for vacuum black holes solutions in $4$-dimensional GR: an asymptotically flat, rotating vacuum black hole
in $4$-dimensional GR is uniquely described in terms of its mass $M$ and angular momentum $J$ \cite{Heusler:1998ua,Chrusciel:2012jk} and is given by the Kerr solution \cite{kerr}. Electromagnetic fields can be introduced into this framework without problem, such that the conjecture can be extended to include the
charge $Q$ of the black hole and the solution be given by the Kerr-Newman black hole \cite{Newman:1965my}, which in the limit of vanishing angular momentum is equal to the unique static, spherically symmetric, charged solution --
the Reissner-Nordstr\"om (RN) solution \cite{reissner}. 
Ever since it had been formulated, the No-hair conjecture has been investigated with respect to extensions of the electro-vacuum, asymptotically
flat case in 4-dimensional GR and it has been demonstrated by explicit construction of solutions that it does not hold
necessarily in the presence of non-linear matter sources such as Skyrme fields \cite{moss} or Yang-Mills fields \cite{bizon}.

The Kerr and RN solutions have played an important r\^ole in the development of Black Hole Thermodynamics \cite{hawking, entropy, bh_thermo}, in particular because of the fact that -- in contrast to the Schwarzschild solution --
they possess an extremal limit with vanishing Hawking temperature. It is exactly an extremal 5-dimensional
black hole solution for which the proportionality between the horizon area and the entropy has been
confirmed by counting its microstates \cite{strominger_vafa}. Due to the existence of this limit and the similarity
of the causal structure, RN solutions are often used as toy models to understand the behaviour of the Kerr solution (see e.g.  \cite{relation_RN_Kerr}). It is interesting to note, however, that the situation changes when adding scalar fields
to these systems. Scalar fields appear in different physical setting, e.g.  in the low energy limits 
of String Theory (e.g. dilaton, axion), but are also used as effective description of physical systems, e.g. in the Ginzburg-Landau model of superconductivity \cite{GL}. Moreover, it has been confirmed now that the Higgs field, a fundamental scalar field, exists in nature \cite{cern_lhc}. Consequently, scalar fields have come back to
the center of attention in black hole physics, although a number of No-hair theorems for scalar fields exist (see e.g. 
\cite{Heusler:1992ss,Bekenstein:1995un,Sudarsky:1995zg,Zannias:1994jf}). 
The requirements used in these latter cases can be evaded, however, when assuming the massive scalar field to possess a 
harmonic time-dependence. This is related to the observation that if the real part of the frequency of this time dependence
is smaller than $m \Omega$, where $m$ is an integer and $\Omega$ is the horizon velocity of the Kerr black hole,
an impinging scalar field can be amplified by the black hole \cite{Bardeen:1972fi,Starobinsky:1973aij,Press:1973zz, Brito:2015oca}. When, additionally, the scalar field is massive, a bounding potential exists that traps these
modes and makes the Kerr black hole superradiantly unstable \cite{Press:1972zz}. 
Furthermore, if the real frequency is just equal to $m\Omega$ (with the imaginary part of the frequency vanishing),
the Kerr black hole can be surrounded by a scalar cloud and when allowing for backreaction between the geometry
and the scalar field, rotating black holes with scalar hair can be constructed \cite{Herdeiro:2014goa,Herdeiro:2014jaa}.
Interestingly, these hairy black holes interpolate between the
standard Kerr black hole and a solitonic-like object without horizon that exists for a complex valued, massive scalar field coupled
minimally to gravity: boson stars \cite{kaup,flp,jetzler,Liddle:1993ha,new2,misch,new1,Kleihaus:2005me,Kleihaus:2007vk}. Note that the rotation is crucial in this case~: hairy black hole analogues of uncharged, spherically symmetric
boson stars do {\it not} exist \cite{Pena:1997cy}. 

For RN black holes a similar process exists~: the amplification of an impinging charged scalar field if the real
part of the frequency is smaller than $q V(r_h)$, where $q$ is the $U(1)$ charge of the scalar field and $V(r_h)$ the
electric potential on the black hole horizon \cite{Bekenstein:1973mi}. The correspondence between the Kerr and
RN solutions might then suggest that a massive scalar field could lead to a superradiant instability for the
RN solution. This assumption is, however, false~: for $M/Q \geq \sqrt{9/8}$ the RN solution is perfectly stable 
\cite{Hod:2013nn}. If the system is put in a cavity, the situation changes and charged black holes with charged scalar hair
can be constructed \cite{Dolan:2015dha}. 

When increasing the number of space-time dimensions, the picture changes. First of all, the No-hair conjecture
is violated for rotating vacuum solutions (see e.g \cite{Emparan:2008eg} and references therein).  In 5 dimensions, e.g., the higher dimensional generalization of the Kerr solution, the Myers-Perry (MP) solution \cite{Myers:1986un} with spherical horizon topology $S^3$ exists, but also two black ring solutions with horizon topology $S^2\times S^1$  \cite{Emparan:2001wn} for the same values of the ADM mass $M$ and the angular momentum $J$. 
The static, spherically symmetric electro-vacuum solutions, on the other hand, are uniquely given by the
Tangherlini-Schwarzschild and Tangherlini-RN solutions \cite{tangherlini}. 
When adding scalar fields with harmonic time-dependence to these solutions, spherically symmetric, non-rotating
boson stars can be constructed with a single complex scalar field \cite{Prikas:2004fx}. However, the existence of hairy generalizations of the MP solution requires a complex scalar field doublet \cite{brihaye_radu_herdeiro}. That it is necessary
to use a doublet in order to obtain rotating solutions with scalar fields in 5 dimensions was realized before in the construction of boson stars \cite{Hartmann:2010pm}. The hairy black holes constructed in \cite{brihaye_radu_herdeiro}
then tend to these solutions in the limit of vanishing horizon radius (see also Section II below for a more detailed explanation), however, are not connected to the MP solutions.

The question then remains what happens to the Tangherlini-RN solutions in higher dimensions with respect to the
formation of charged scalar hair. In the Appendix of this paper, we show that the dimensionality does not influence the
fact that spherically symmetric, asymptotically flat electro-vacuum black holes do not support charged scalar fields
with harmonic time-dependence. 

Higher dimensional gravity models can be extended beyond the Einstein gravity case. A hierarchy of
higher order curvature corrections, of which the lowest order term is the Gauss-Bonnet (GB) term, can be added
and render the equations of motion still 2nd order \cite{lovelock}. The GB term contains suitable quadratic combinations
of the Riemann and Ricci tensors as well as the Ricci scalar and is a total divergence in 4 space-time dimensions. 
Vacuum \cite{Boulware:1985wk} and electro-vacuum \cite{Wiltshire:1985us} solutions to the corresponding equations of motion have been found. Extending the model to include scalar fields,
non-rotating and rotating uncharged boson stars in 5 dimensional Einstein-Gauss-Bonnet (EGB) can be  constructed,  \cite{Hartmann:2013tca} and \cite{Brihaye:2013zha}, respectively. Note that in the later case, again a complex doublet of scalar fields has been used. Rotating EGB black holes with uncharged scalar hair were shown to exist in \cite{Brihaye:2015qtu}.
An interesting new feature appearing in 5-dimensional EGB gravity is that now static, electrically charged black hole
solutions with charged scalar hair (that possesses harmonic time-dependence) do exist \cite{Grandi:2017zgz}.
For these solutions, the real-valued frequency of the time-dependence has to be chosen equal to the upper bound on the
superradiant frequency in analogy to the construction of hairy, rotating Kerr black holes in 4 dimensions \cite{Herdeiro:2014goa,Herdeiro:2014jaa}. The authors of \cite{Grandi:2017zgz} also studied the corresponding boson star
solutions, however, did not discuss the critical phenomena of the solutions and if and how the hairy black holes
are connected to the boson star solutions and/or the branches of non-hairy, charged black holes that exist in this model.
The aim of this paper is to answer these questions and also provide a simple argument for the existence
of an upper bound on the $U(1)$ charge of the scalar field. While the focus of the analysis in \cite{Grandi:2017zgz}
was the determination of the domain of existence in relation to the charge of the scalar field $q$ and the GB coupling 
$\alpha$, we study the pattern mainly in function of the frequency of the scalar field's time-dependence, $\omega$,
and the charge $Q$ of the black holes. 

Our paper is organized as follows~:  in Section II, we give the model, Ansatz and boundary conditions
for the different solutions available. In Section III, we discuss the black hole solutions, first the charged EGB 
black holes carrying scalar field clouds on their horizon, and then the charged, hairy black hole solutions, while
Section IV contains the discussion of the globally regular solutions, charged boson stars. 
We conclude in Section V.

\section{The model and Ansatz}
We want to study charged  black holes and boson stars
in 5-dimensional Einstein-Gauss-Bonnet (EGB) gravity.
The model is described by the following action~:
\begin{equation}
\label{egbbs}
   S = \int {\rm d}^5 x \sqrt{- g} \left[ \frac{R}{16\pi G_5}  + \alpha {\cal L}_{\rm GB} 
   -  D_M \Phi^{*} D^M \Phi - m^2 \Phi^{*} \Phi - \frac{1}{4} F^{MN} F_{MN} \right]    \  , \ \ \ M,N  \in \{0,1,2,3,4 \}     \ .
\end{equation}
Here $R$ represents the Ricci scalar, $G_5$ is the $5$-dimensional Newton's constant, $\alpha$ is the Gauss-Bonnet (GB) coupling
and ${\cal L}_{\rm GB}$ is the Gauss-Bonnet Lagrangian density given by~:
\begin{equation}
        {\cal L}_{\rm GB} = R^{MNKL} R_{MNKL} - 4 R^{MN} R_{MN} + R^2 \ \ ,  \ \  M,N,K,L \in \{0,1,2,3,4 \}   \ .
\end{equation}
This second order gravity action is coupled minimally to a complex scalar field $\Phi$ that is gauged under a $U(1)$ symmetry with
gauge field $A_M$ and field strength tensor $F_{MN}=\partial_M A_N  - \partial_N A_M$, $M,N \in \{0,1,2,3,4 \}$, while
the covariant derivative operator reads $D_M = \nabla_M - i q A_M$, $M \in \{0,1,2,3,4 \}$. $q$ denotes the $U(1)$ charge of the scalar field.
Variation of the action (\ref{egbbs}) with respect to the metric, the gauge field and the scalar field, respectively, leads to a set of coupled, non-linear differential
equations. 

In the following, we will use the following spherically symmetric, static Ansatz for the metric~:
\begin{equation}
      ds^2 = - b(r) dt^2 + \frac{1}{f(r)} dr^2 + r^2 d \Omega_3^2 \ ,
\label{ss_metric}
\end{equation}
where $d\Omega_3^2$ denotes the line element of the unit 3-sphere $S^3$.
For the gauge field and complex scalar field, we choose~:
\begin{equation}
        A_M dx^M = V(r) dt  \ \ \ , \ \ \ \Phi(t,r) = e^{- i \omega t} \phi(r)
\end{equation}
where $\omega$ is a real constant. Note that while the gauge and metric fields are static, the scalar field possesses a harmonic time dependence, although the energy-momentum tensor will be static and spherically symmetric. 
Of course, we could choose a gauge such that $\Phi$ becomes also static (and real), however and as explained in more detail below, we need to have $\omega > 0$ in order to be able to construct charged EGB black holes with scalar hair.
With this Ansatz, we obtain a system of four non-linear, ordinary differential
equations (plus a constraint). The equations for the functions $b(r)$, $V(r)$ and $\phi(r)$ are of second order, while
the equation for $f(r)$ is of first order.  For the explicit form of these equations -- albeit in a slightly different parametrization of the metric -- 
we refer the reader to  \cite{Grandi:2017zgz}.

Let us finally point out that with suitable rescalings of the fields and the radial variable $r$, we can set two of the four couplings in 
the model to fixed values without loss of generality. In the following, we will set $16\pi G_5\equiv 1$ and $m\equiv 1$. 
The two remaining -- appropriately rescaled and hence dimensionless couplings -- are then the GB coupling $\alpha$ and the charge $q$ of the scalar field.
Note that with our rescalings, the couplings in \cite{Grandi:2017zgz} (in the following denoted by a tilde)
are  related to ours according to
\begin{equation}
\phi = \sqrt{2} \tilde{\phi} \ \ , \ \ V = \sqrt{2}\tilde{h}  \ \ , \ \ Q = \sqrt{2} \tilde{Q} \ \ , \ \ q = \frac{1}{\sqrt 2} \tilde{q}
 \ \ .    \label{correspondance_gl}       
\end{equation}
 
As  demonstrated in \cite{Grandi:2017zgz}, the system of coupled equations admits globally regular as well as black hole solutions. In order to 
be able to investigate (primarily) the black hole solutions with respect to the No-hair conjecture, we will further characterize
these solutions by their ADM mass $M$ and charge $Q$. These are physical quantities subject to a Gauss law and can hence be read of from 
the behaviour of the solutions at spatial infinity $r\rightarrow \infty$, i.e. in our choice of parametrization from the behaviour of the
functions $f(r)$ and $V(r)$~:
\begin{equation}
          f(r\rightarrow \infty)\rightarrow 1 - \frac{M}{r^2} + {\rm O}(r^{-3}) \ \ , \ \ 
          V(r \to \infty) \rightarrow V_{\infty} - \frac{Q}{r^2} + {\rm O}(r^{-3})   \ ,
\end{equation}
where $V_{\infty}$ is a constant that we will choose equal to zero in the following. 

\section{Charged boson stars}
As indicated in the Introduction, solitonic-like objects, so-called boson stars exist in self-gravitating scalar field models
with harmonic time-dependence of the scalar field. These play an important r\^ole in the understanding of the pattern of black hole
solutions and we will compare their properties in the following with those of EGB black with and without scalar hair. 
 
In order to find regular solutions on the interval $r\in [0:\infty[$, we need to impose the following boundary conditions at $r=0$~: 
\begin{equation}
            f(0) = 1 \ , \ b'(0) = 0 \ , \  q V(0) + \omega = 0   \    , \ \phi'(0) = 0  \ .
\end{equation}
Boson stars are characterized  by their frequency $\omega$, mass $M$ and charge $Q$.
Charged EGB boson stars exist only for small values of $q$ (in fact for $q^2 < 1/3$ as found in \cite{Grandi:2017zgz})
and on a limited interval of $\omega \in [\omega_{\rm min}, \omega_{\rm max}]$, where $\omega^2_{\rm max}=m^2$ for
the non self-interacing case. Hence, with our choice $m=1$ we have  $\omega_{\rm max}=1$. 
In the limit $\omega\rightarrow 1$, a configuration with finite mass $M$ and $Q$ is approached, for which, however, the
metric function $b(0)= g_{tt}(0)$ tends to zero. This corresponds to a singular solution. 
Increasing $q$,  the interval of  allowed frequencies  decreases tending to zero in the limit $q^2 \to 1/3$.  

\begin{figure}[ht]
\begin{center}
{\includegraphics[width=8cm]{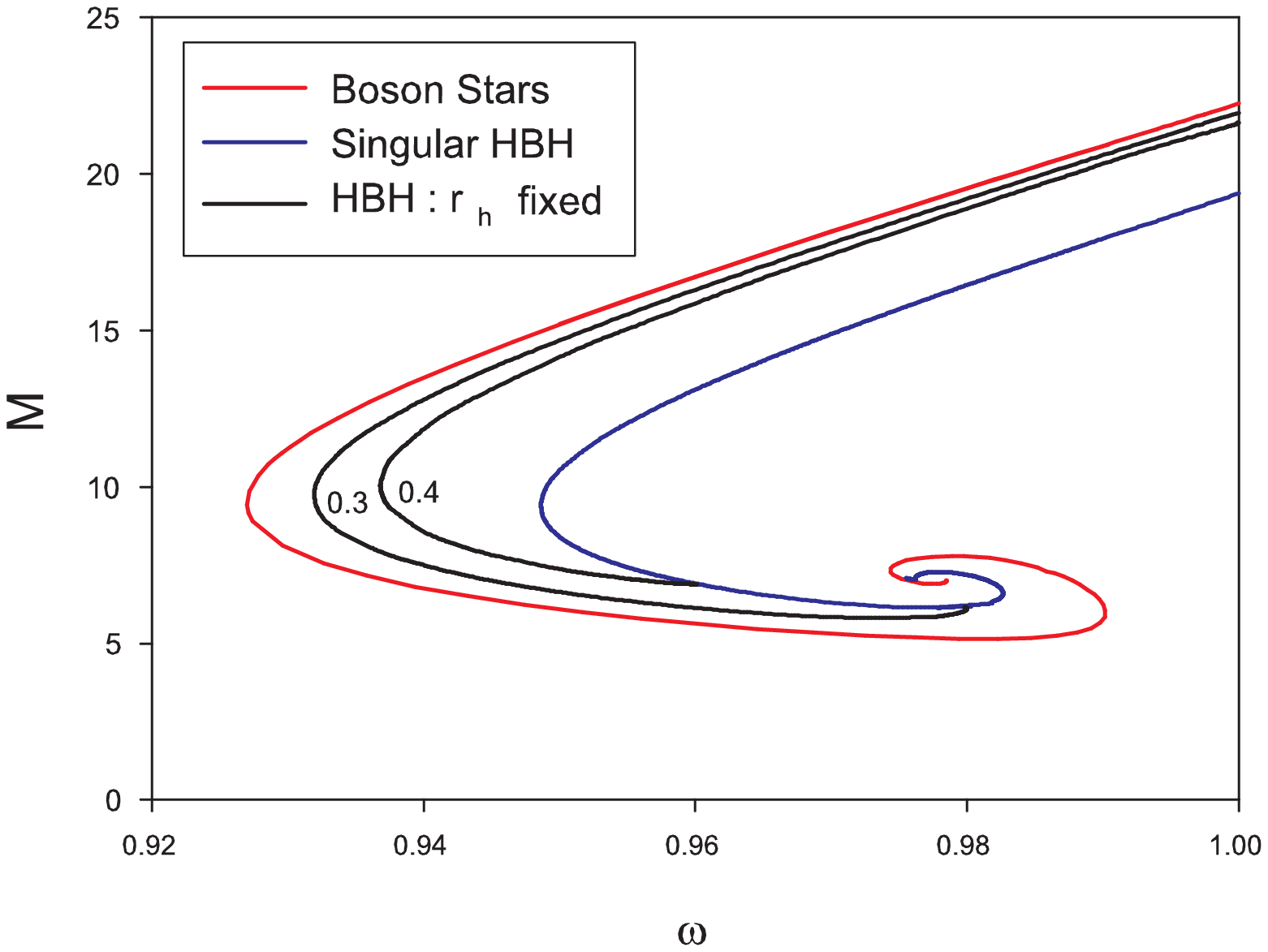}}
{\includegraphics[width=8cm]{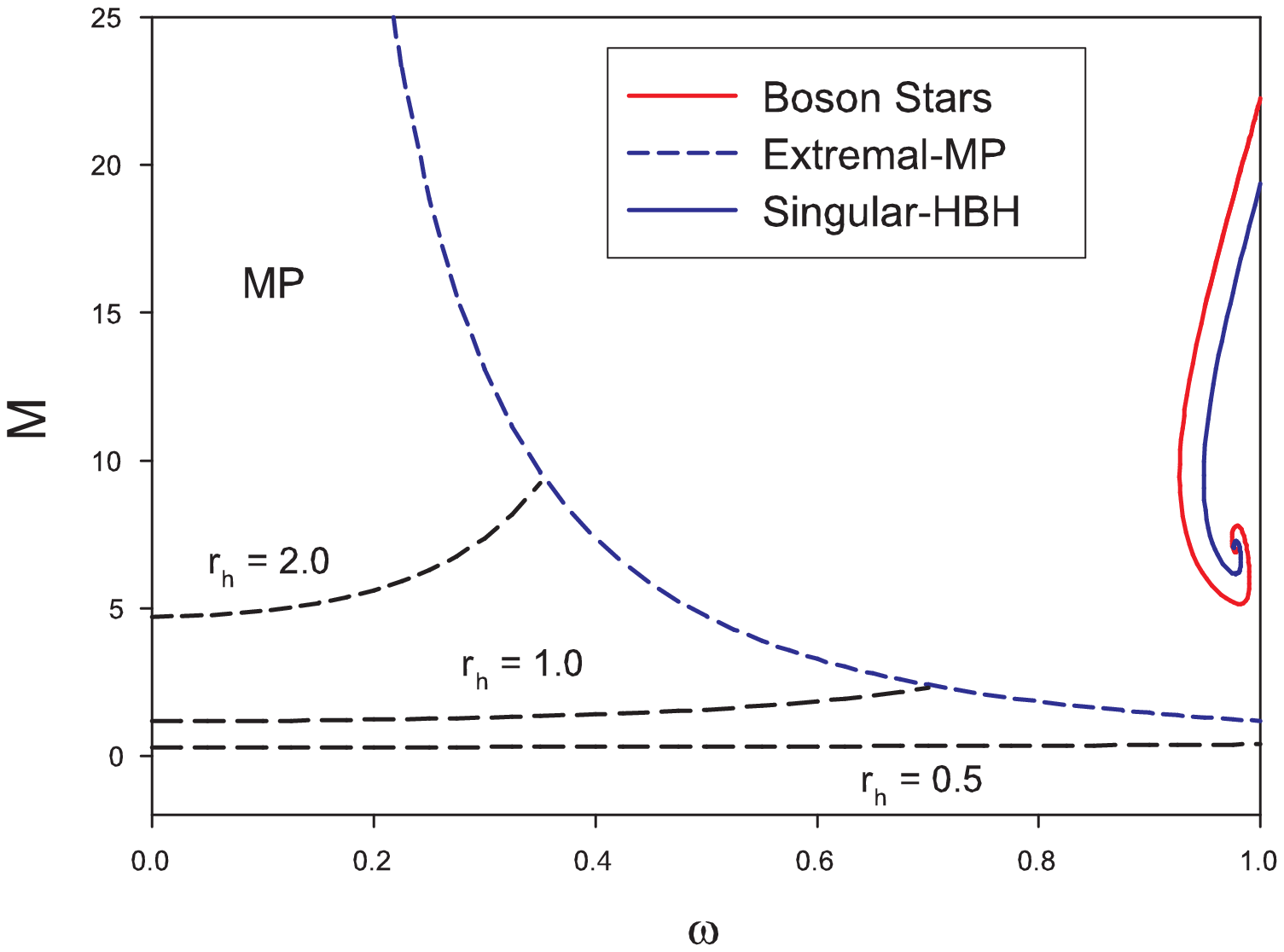}}
\end{center}
\caption{We show the spectrum of rotating, asymptotically flat solutions available in 5-dimensional Einstein gravity minimally coupled to a complex scalar doublet.
Left: We show the dependence of the mass $M$ on the frequency for boson stars (red), for hairy Myers-Perry black holes (HBH) with two different values of the horizon radius, $r_h=0.3$ and
$r_h=0.4$ (black) as well as the singular limit of the HBH, which has vanishing $T_{\rm H}$, but is not
regular at the horizon (blue). 
Right: Same as left, but in addition we show the corresponding Myers-Perry solutions (MP) without scalar hair (dashed),
with non-extremal (black) joining the branch of extremal (blue) black holes.
\label{massomega}
}
\end{figure}

In order to understand the difference between the uncharged rotating boson stars (see Fig.\ref{massomega})
and the charged non-spinning ones considered in this paper, let us point out that (as we will demonstrate in the following section)
the branches of boson stars remain disjoined from the hairy charged, static EGB black holes and do {\it not} 
appear in the limit of vanishing horizon radius.
In particular, our numerical results indicate that the
metric function $f(r)$ approaches a step function at $r=r_h$ for $r_h\rightarrow 0$.  Accordingly, the Hawking temperature $\sim f'(r_h)$ (see (\ref{temperature}) below) increases.
For all  hairy, charged EGB black holes that we constructed, our numerical results
strongly suggest that, in the limit $r_h \to 0$, solutions with $Q=0$ are approached such that 
in agreement  with (\ref{temperature}) the Hawking temperature remains finite.

\section{Black holes} 
For $\phi(r)\equiv 0$, the model (\ref{egbbs}) admits a two-parameter family of closed form solutions 
\cite{Wiltshire:1985us}~:
\begin{equation}
\label{egbbh}
     f(r) = b(r) = 1 + \frac{r^2}{4 \alpha} \left( 1 - \sqrt{1 + 4 \alpha \left(\frac{2M}{r^4} - \frac{2 Q^2}{3 r^6}\right) } \right)
      \ \ ,  \ \ V(r) = - \frac{Q}{r^2}   \ .
\end{equation}
The condition $f(r_h)=0$ imposed on the outer horizon $r=r_h$ relates the mass $M$ and charge $Q$ of these charged Einstein-Gauss-Bonnet (EGB) black hole solutions to each other~:
\begin{equation}
\label{horizon}
r_{h}^2 = \frac{M}{2}-\alpha + \sqrt{\frac{(M-2\alpha)^2}{4}-\frac{Q^2}{3}} \ \ , \ \ {\rm resp.}   \ \ \ M=2\alpha + \frac{Q^2}{3r_h^2} + r_h^2  \ .
\end{equation}  
Note that the condition $f(r)=0$ has two possible solutions: $f(r_h)=0$ at the outer horizon, but also
$f(r_{-})=0$, where
\begin{equation}
r_{-}^2= \frac{M}{2}-\alpha - \sqrt{\frac{(M-2\alpha)^2}{4}-\frac{Q^2}{3}}
\end{equation}
The Hawking temperature $T_{H}$ and entropy $S$, respectively, of these black holes is given by~:
\begin{equation}
\label{temperature}
             T_H = \frac{f'(r_h)}{4 \pi}  = \frac{1}{4 \pi} \frac{2(3 r_h^4 - Q^2)}{3 r_h^3(4\alpha + r_h^2)} \ \  , \ \  S=\int\limits_0^{r_h} T_{\rm H}^{-1}\left(\frac{\partial M}{\partial r_h}\right) {\rm d} r_h =  \frac{4\pi}{3}\left(r_h^3 + 12\alpha r_h\right)  \  \ .
\end{equation}
As is evident from (\ref{temperature}), an extremal limit with $T_{\rm H}=0$ and finite entropy $S$
exists. In this case, the mass $M$ and the charge $Q$ read
\begin{equation}
M_{\rm ex}= \frac{2}{\sqrt{3}} Q_{\rm ex} + 2\alpha = 2(r_{h,{\rm ex}}^2 + \alpha)  \ \ , \ \ 
r_{h,{\rm ex}} = \sqrt{\frac{Q_{\rm ex}}{\sqrt{3}}}   \ .
\end{equation}
For $\alpha=0$ these black hole solutions are the
$5$-dimensional generalizations of the Reissner-Nordstr\"om solutions \cite{tangherlini} with $f(r)=1-\frac{M}{r^2} + \frac{Q^2}{3 r^4}$. 
For the latter case, we show in the Appendix that these solutions cannot carry scalar hair. On the other hand, this is not true for the case $\alpha > 0$ \cite{Grandi:2017zgz}~: for sufficiently large charge $q$ of the scalar field, the charged EGB black holes become unstable to scalar hair formation. 

\subsection{A short note on Myers-Perry (MP) black holes with scalar hair}
In order to be able to compare the different branches of globally regular and black hole solutions 
available in 5 dimensions when scalar fields are present, we would like to make this short note
on MP solutions, i.e. uncharged, rotating black holes in 5 dimensions, with scalar hair as well as the corresponding
boson star solutions in Einstein gravity. These solutions have been discussed in \cite{Brihaye:2014nba} and
we refer the reader for more details to this paper. Here we present two plots in Fig.~\ref{massomega} not previously published that we believe to be instructive. 
The left side of Fig.~\ref{massomega} shows the dependence of the mass $M$ of the solutions
on the frequency $\omega$. As is evident from this plot, the boson stars appear as a regular limit ($r_h\rightarrow 0$) of the hairy Myers-Perry solutions (HBH), for which we give the data for two different horizon values $r_h=0.3$ and
$r_h=0.4$, respectively. In particular, in this limit the temperature (resp. area) of the HBH tends to infinity (resp. zero). On the other hand, at a minimal value of the mass $M$ these branches of HBH join a branch
of singular black hole solutions, i.e. solutions with vanishing Hawking temperature $T_{\rm H}$ that do not fulfil the
regularity conditions for the scalar field.  Close to the maximal frequency (i.e. $\omega \sim 1$) the HBH exist for $r_h \in [0, 0.7]$. On the right hand side of Fig.~\ref{massomega} we show these exact branches of solutions together with
the corresponding non-hairy MP solutions. This plot clearly demonstrates that the HBH  and boson star branch, respectively, are disjoined from the branches of MP solutions, in contrast to the 4-dimensional case \cite{Herdeiro:2014goa,Herdeiro:2014jaa}.

\subsection{Charged EGB black holes surrounded by charged scalar field clouds}
In order to understand the behaviour of the charged EGB black holes and their instability with respect to scalar hair formation, we follow \cite{Grandi:2017zgz} and first 
investigate the case of a scalar field in a fixed black hole background space-time of the
form (\ref{egbbh}).  We hence solve the scalar field equation on the interval $r\in [r_h:\infty[$ and impose the regularity of the scalar field on the horizon $r=r_h$.  This leads
to the condition $\omega = q Q /r_h^2$ on the angular frequency. 
The relevant equation for $\phi(r)$ then reads~:
\begin{equation}
\phi'' = - \left(\frac{3}{r} + \frac{f'}{f}\right) \phi' 
+ \left(\frac{m^2}{f} - \frac{q^2 Q^2}{f^2} \left(\frac{1}{r^2} - \frac{1}{r_h^2} \right)^2 \right) \phi \ \ , 
\label{klein_gordon}
\end{equation}
where the prime now and in the following denotes the derivative with respect to $r$.
Note that since $Q$ and $r_h$ are fixed by the background solution, the gauge coupling $q$ becomes the spectral parameter of this equation. In other words, for fixed $Q$ and $ r_h$, normalizable solutions of (\ref{klein_gordon}) will exist only for 
discrete values of $q$. Here, we will only study the fundamental mode,
 characterized by a profile of $\phi(r)$ without nodes, but we conjecture that
solutions with a number of node exist in this model. 

 The equation (\ref{klein_gordon}) has to be solved numerically subject to the following boundary conditions~: 
 \begin{equation}
       \phi(r_h) = 1 \ \ , \ \ \phi'(r_h) = \frac{m^2}{f'(r_h)} \phi(r_h) \ \  \ \ 
       {\rm with }\ \ f'(r_h) = \frac{2(3 r_h^4 - Q^2)}{3 r_h^3(\alpha + r_h^2)} \  
 \end{equation}
 at $r=r_h$ and $\phi(r\to\infty)\to 0$, respectively. 
Note that the behaviour of $f'(r_h)$ fixes an upper limit on the charge $Q$ 
 given by  $Q^2 \leq 3 r_h^4$, which comes from the requirement of positive Hawking temperature $T_{\rm H}$ (see (\ref{temperature}). 
The condition on $\phi(r_h)$ fixes the normalisation of the scalar field, the condition on $\phi'(r_h)$ is the regularity condition, while the behaviour of $\phi$ at spatial infinity comes from the requirement of finite energy. Since $\phi(r)$ should be exponentially
 decaying at infinity, we also have a bound on the frequency $\omega$ which is $\omega^2 - m^2  < 0$ and hence
 $q^2 Q^2 \leq m^2 r_h^4$.  Combining these two constraints on $Q$, using that the maximally charged, extremal
 black hole solution has $Q^2=3r_h^4$ we find that $q^2 \leq q^2_{\rm cr}= m^2/3$, which is the bound on $q$ found numerically in \cite{Grandi:2017zgz}.

\begin{figure}[ht]
\begin{center}
{\includegraphics[width=8cm]{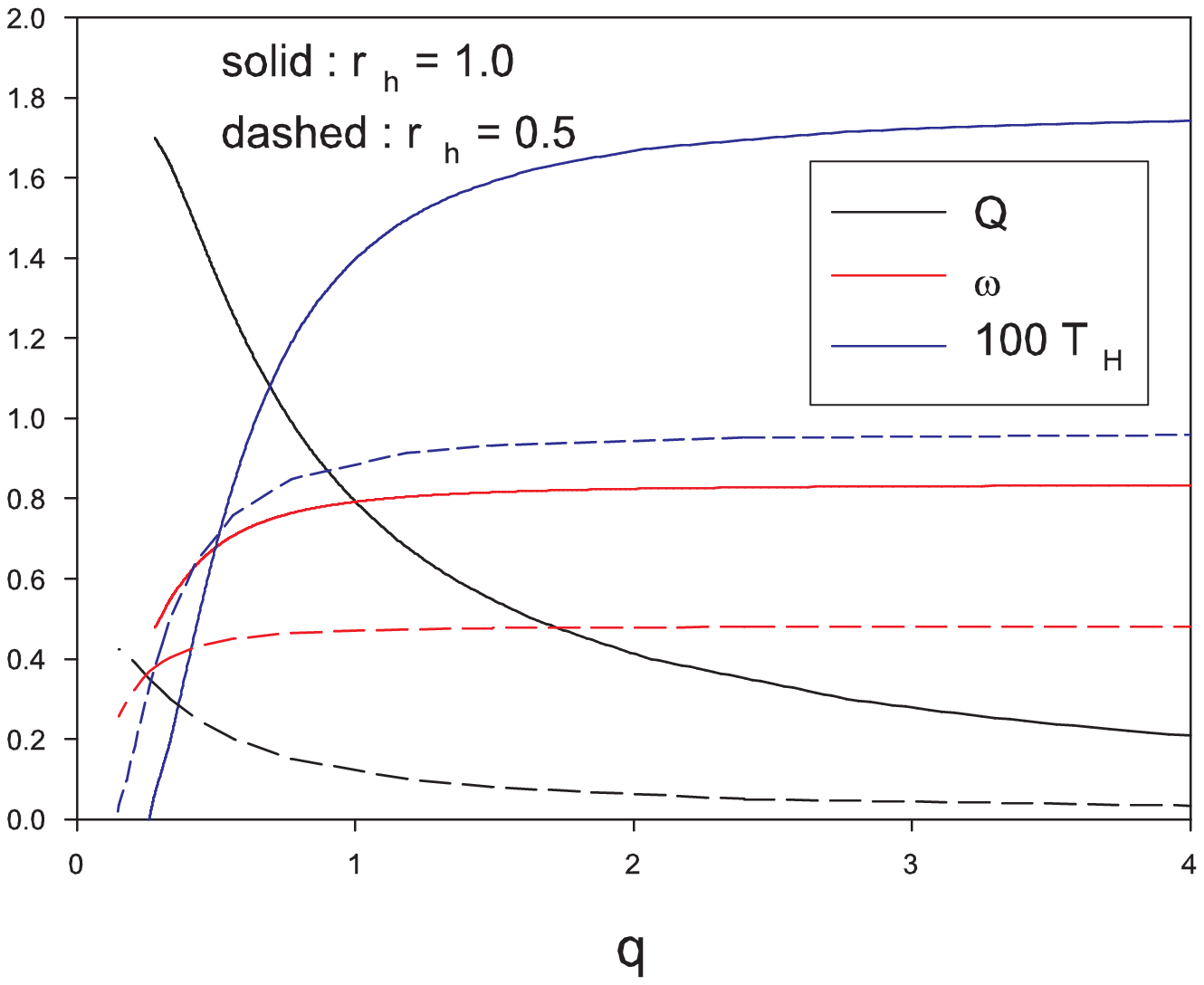}}
{\includegraphics[width=8cm]{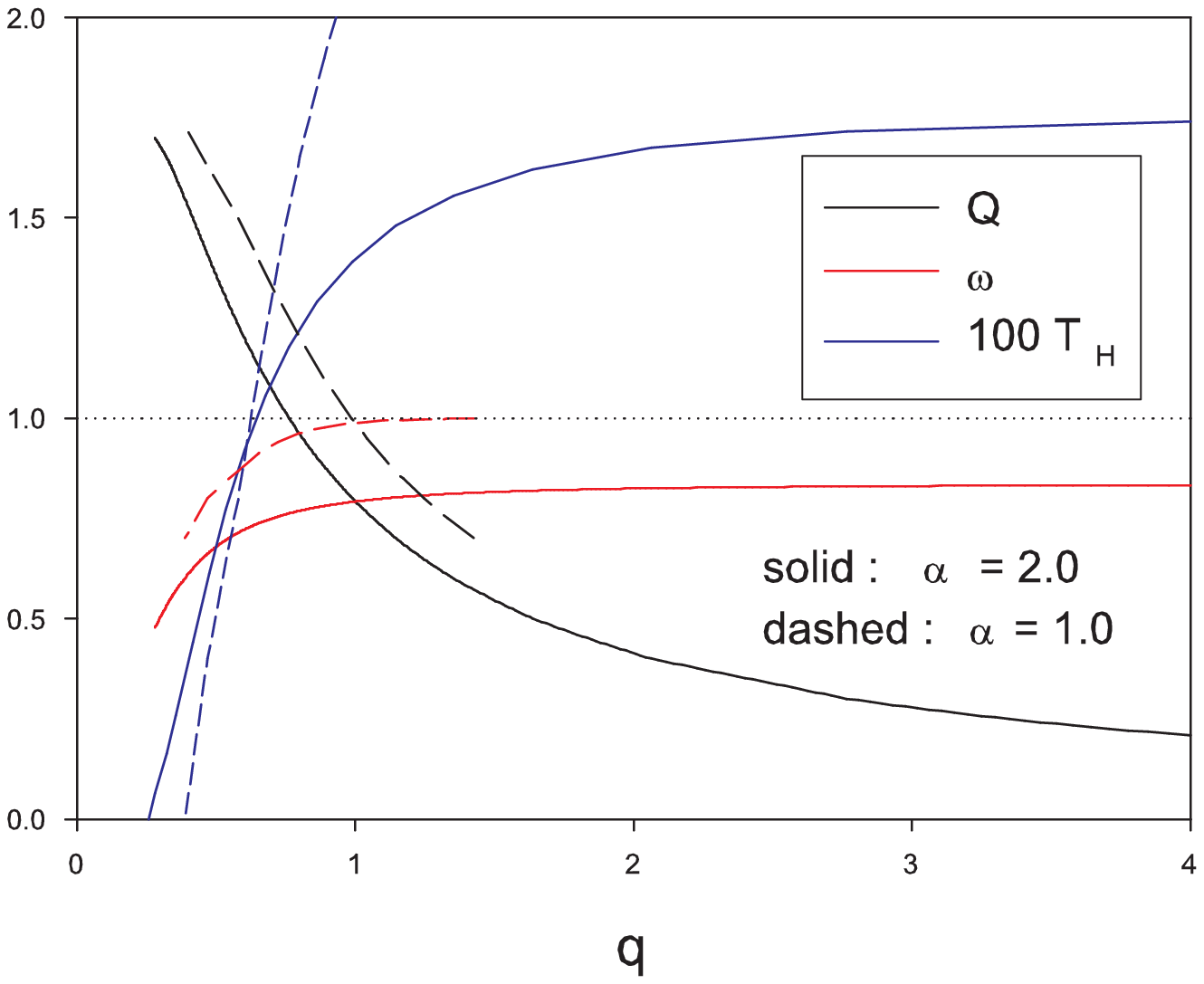}}
\end{center}
\caption{Left: We give the charge $Q$ (black), the frequency $\omega =q Q /r_h^2$ (red) and the Hawking temperature $T_{\rm H}$ (blue) of EGB black holes with $\alpha=2.0$ and two different values of the horizon 
radius, $r_h=0.5$ (dashed) and $r_h=1.0$ (solid), surrounded by a charged scalar field cloud
in dependence of the charge $q$ of the scalar field.
Right: Same as left, but for $r_h=1$ fixed  and two different values of $\alpha=1.0$ (dashed) and $\alpha=2.0$ (solid), respectively. \label{clouds}}
\end{figure}

We have solved the equation (\ref{klein_gordon}) numerically using a collocation solver
with adaptive grid selection \cite{COLSYS}. We have fixed the parameters $\alpha$ and $r_h$ and
constructed branches of EGB black holes with charge $Q$ surrounded by clouds of scalar fields with $U(1)$ charge $q$.
Our results are shown in Fig.~\ref{clouds}, where 
we give the Hawking temperature $T_{\rm H}$ (blue curves), the frequency $\omega$ (red curves) as well as the charge $Q$ (black curves), for which charged scalar
clouds with fixed $q$ exist, in function of $q$. The left part of Fig.~\ref{clouds} shows the data for two
black holes with different horizon radii, $r_h=0.5$ (dashed curves) and $r_h=1.0$ (solid curves), respectively, for fixed value of $\alpha$, while
on the right side of Fig.~\ref{clouds}, we give the data for fixed $r_h=1.0$ and $\alpha=1.0$ (dashed curves), respectively $\alpha=2.0$ (solid curves).  
We observe that solutions with charged scalar clouds exist only in a finite interval of $q$, i.e. $q\in [q_{\rm min}:q_{\rm max}]$. The lower limit  of the interval, $q_{\rm min}$,  corresponds to the maximal possible value of $Q$, i.e.
to the  extremal limit of the EGB background black hole, which is reached for $Q = \sqrt{3} r_h^2$. This is clearly indicated by the fact that $T_{\rm H}\rightarrow 0$ in this limit. When increasing $q$ we expect that the corresponding
$Q$ should decrease in order to avoid too strong electric repulsion. These is indeed what we observe.  Decreasing
$Q$ means moving away from the extremal limit and $q$ is only limited by the requirement  $\omega=qQ/r_h^2 \leq 1$. 
This is clearly visible for $r_h=1$ and $\alpha=1$ in the right side of Fig.~\ref{clouds}.
To summarize, we observe that the interval in $q$ for which charged scalar clouds exist, decreases with increasing
horizon radius $r_h$ and decreasing $\alpha$, respectively. In the limit $\alpha\rightarrow 0$, the interval
should shrink to zero as no charged scalar clouds exist on Tangherlini-RN solutions (see Appendix). 

Let us finally stress that for fixed value of $\alpha$, $q$, and $r_h$ we would expect the corresponding values of $Q$ and $\omega$ to correspond 
to the values from where the hairy EGB black holes bifurcate from the family of EGB black holes given in (\ref{egbbh}).
This is, indeed, the case, as we will discuss in the following subsection.

\begin{figure}[ht]
\begin{center}
{\includegraphics[width=8cm]{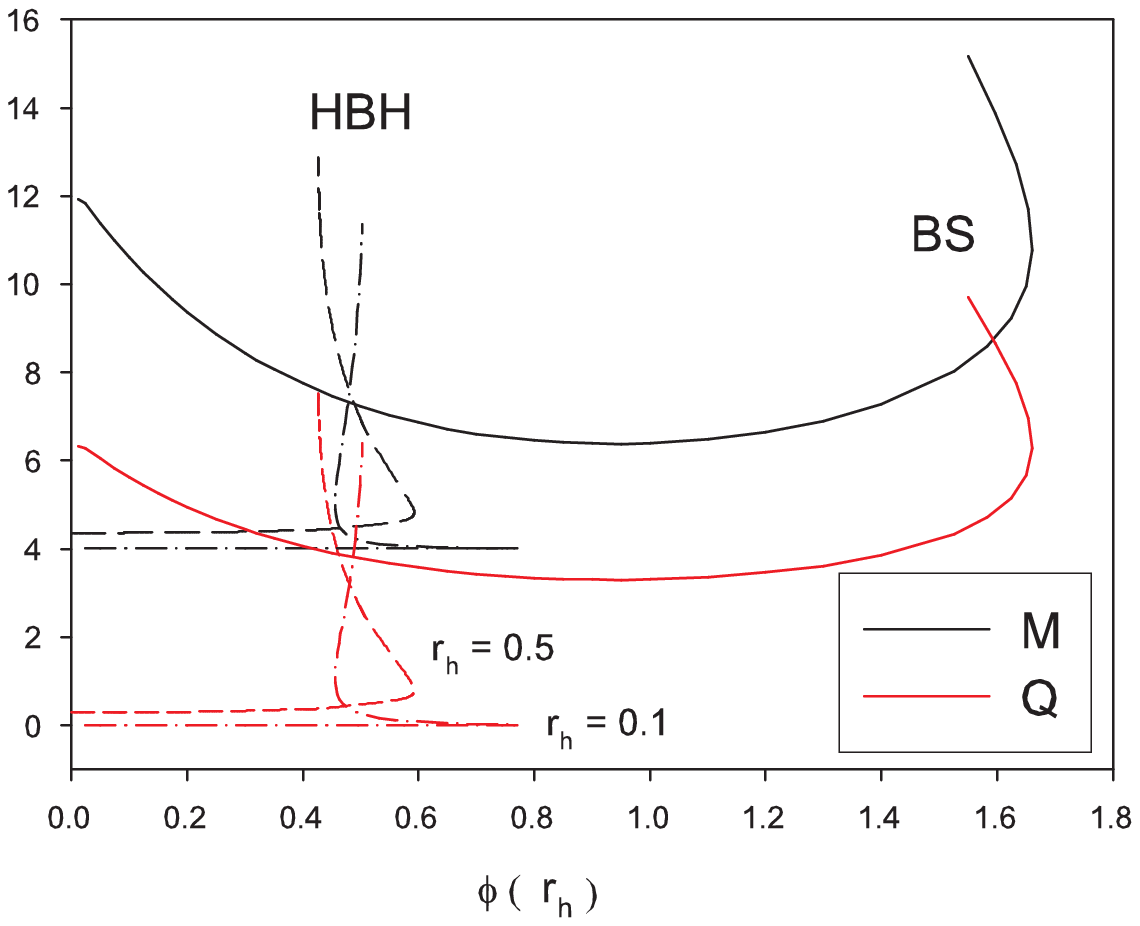}}
{\includegraphics[width=8cm]{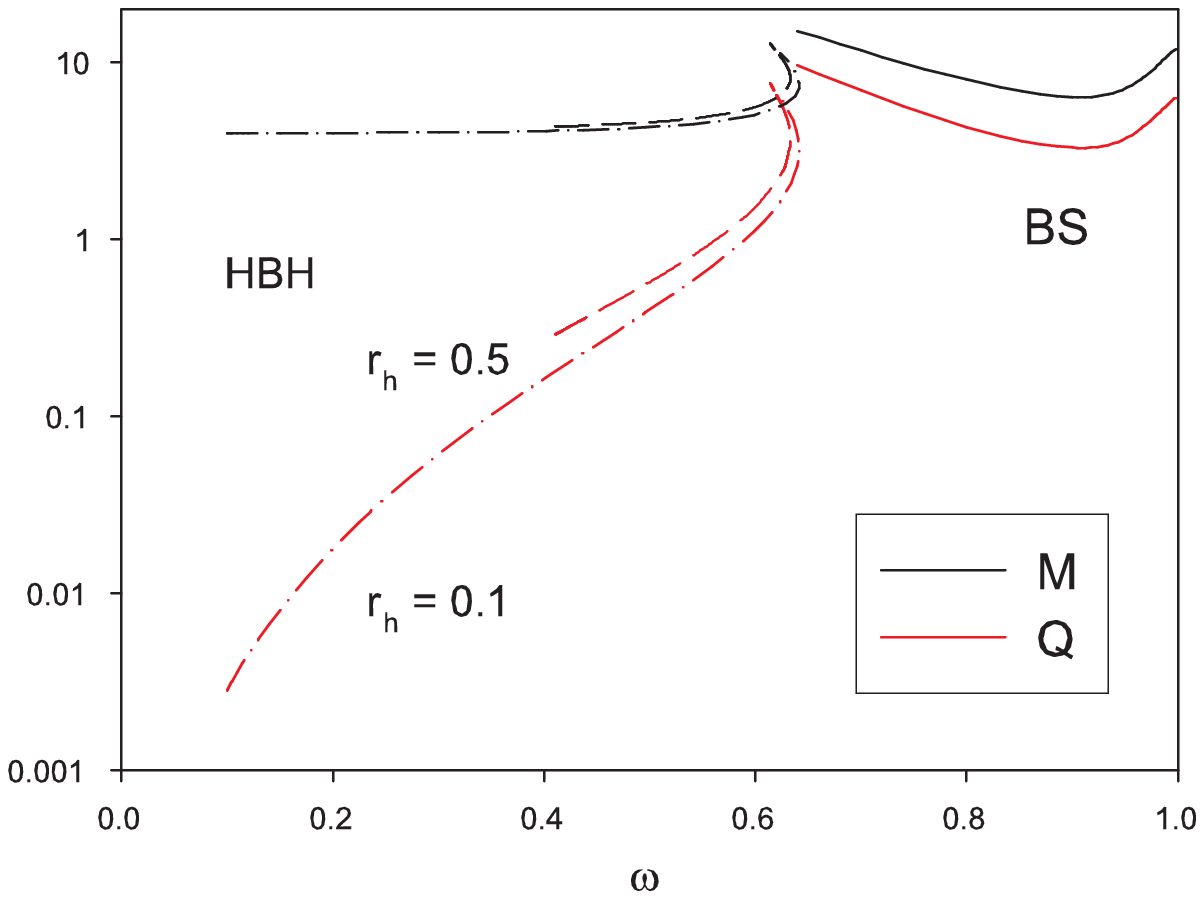}}
\end{center}
\caption{We show the properties of hairy EGB black holes (HBH) as well as of the corresponding boson stars (BS) for 
$\alpha=2.0$ and $q=0.3535$. 
Left: The mass $M$ (black) and the charge $Q$ (red) of HBH with $r_h=0.1$ and $r_h=0.5$, respectively, as function of
the value of the scalar field on the horizon $\phi(r_h)$ (dashed) as well as for BS as function of the scalar field at the origin $\phi(0)$ (solid). 
 Right: Same as left, but 
as a function of the frequency $\omega$.
\label{q_035}
}
\end{figure}

\subsection{Black holes with scalar hair}
Given the values of the parameters for which charged scalar clouds on EGB black hole exist, a
family of hairy black holes (HBH) can be constructed by imposing the following regularity conditions on the horizon $r=r_h$~:
\begin{equation}
\label{regular_horizon}
     f(r_h)=0 \ , \ b'(r_h) = 0 \ , \ q V(r_h) + \omega = 0 \ , \ \phi'(r_h) = \frac{m^2}{f'(r_h)}\phi(r_h) \  , \ 
\end{equation}
 where 
\begin{equation}
\label{regular_horizon2}
 f'(r_h) = \left(2 r b' \frac{6 - \kappa r^2 m^2 \phi^2}{6 b'(4\alpha+r^2)+ r^3 (V')^2}\right)_{r=r_h}  \ .
\end{equation}
The condition (\ref{regular_horizon2}) relates the frequency $\omega$ of the scalar field to the value of the electric
potential on the horizon $V(r_h)$. Note that this is the counterpart of the synchronization condition used in the
construction of hairy rotating black holes  (see e.g. \cite{Herdeiro:2014goa},\cite{Brihaye:2014nba}). 

We require the solutions to be asymptotically flat, hence impose the following conditions on the functions ~:
\begin{equation}
\label{asymptotic}
       b(r\to \infty) = 1 \ \  , \ \ V(r \to \infty)  = \frac{Q}{r^2} \ \ , \ \ \phi(r \to \infty) = 0   \ .
\end{equation}
\begin{figure}[ht]
\begin{center}
{\includegraphics[width=8cm]{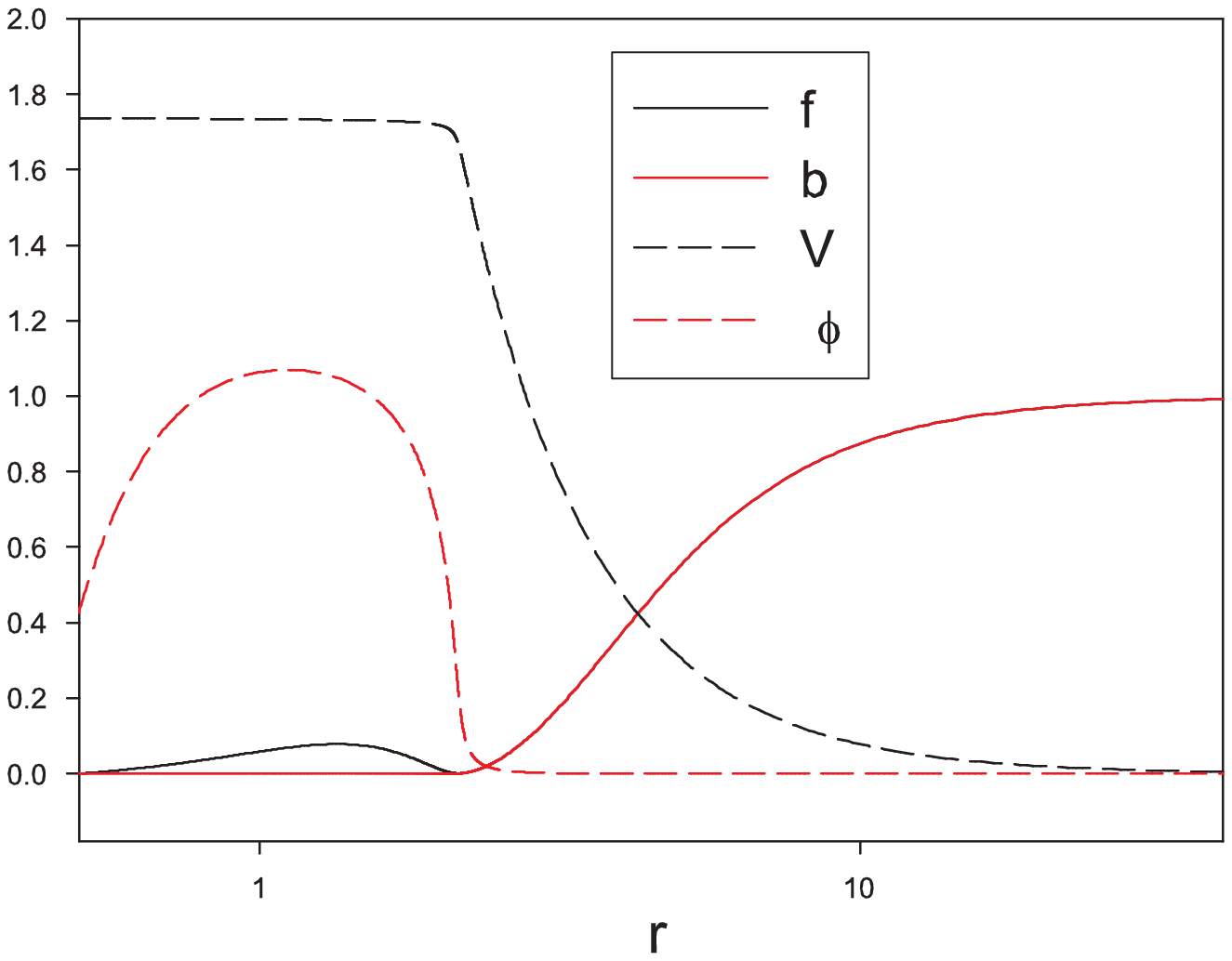}}
{\includegraphics[width=8cm]{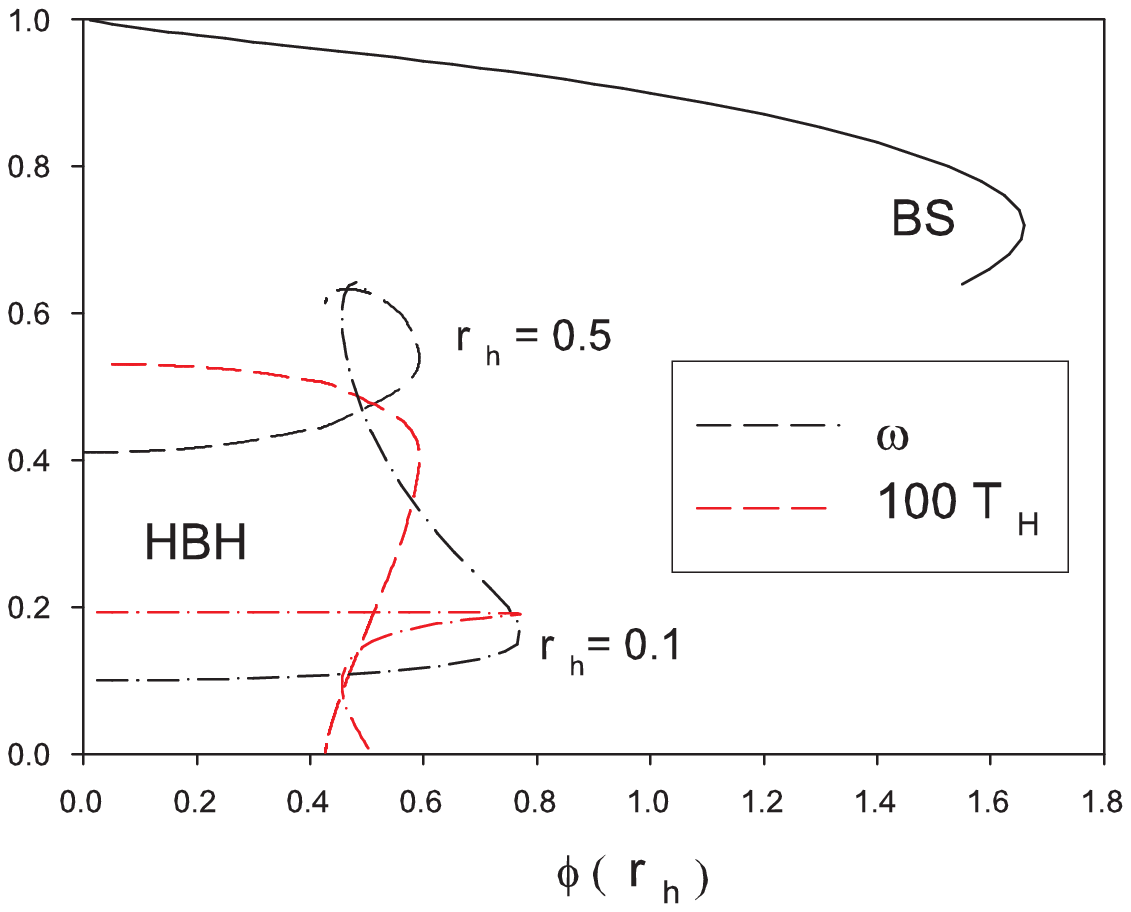}}
\end{center}
\caption{Left: We show the metric functions $f(r)$ and $b(r)$ as well as the matter functions
$V(r)$ and $\phi(r)$ for a hairy EGB black hole (HBH) with $\alpha=2.0$, $q=0.3535$ and $r_h=0.5$
close to the formation of an extremal solution.
Right: We show the dependence of the frequency $\omega$ (black dashed) and the Hawking temperature $T_H$ (red dashed) on $\phi(r_h)$ for the HBH solutions in Fig.\ref{q_035}. We also give $\omega$ as function of $\phi(0)$ for the boson star (BS) solutions (solid). 
\label{q_035b}
}
\end{figure}  
Because of the big number of parameters in the model, we discuss in the following the case $\alpha=2.0$ in more detail.
In the following, we will denote by $\omega_{\rm cl}$ the value of $\omega$ for which the scalar field cloud exists on
the EGB black holes. Hairy EGB will emerge from the corresponding cloud for $\omega > \omega_{\rm cl}$. 
In order to discuss the pattern of solutions, we can then subdivide the interval of $q$ into three parts using the critical value of $q_{\rm cr}=m/\sqrt{3}\equiv 1/\sqrt{3}\approx 0.57745$ existing for the scalar clouds. 

\begin{figure}[ht]
\begin{center}
{\includegraphics[width=8cm]{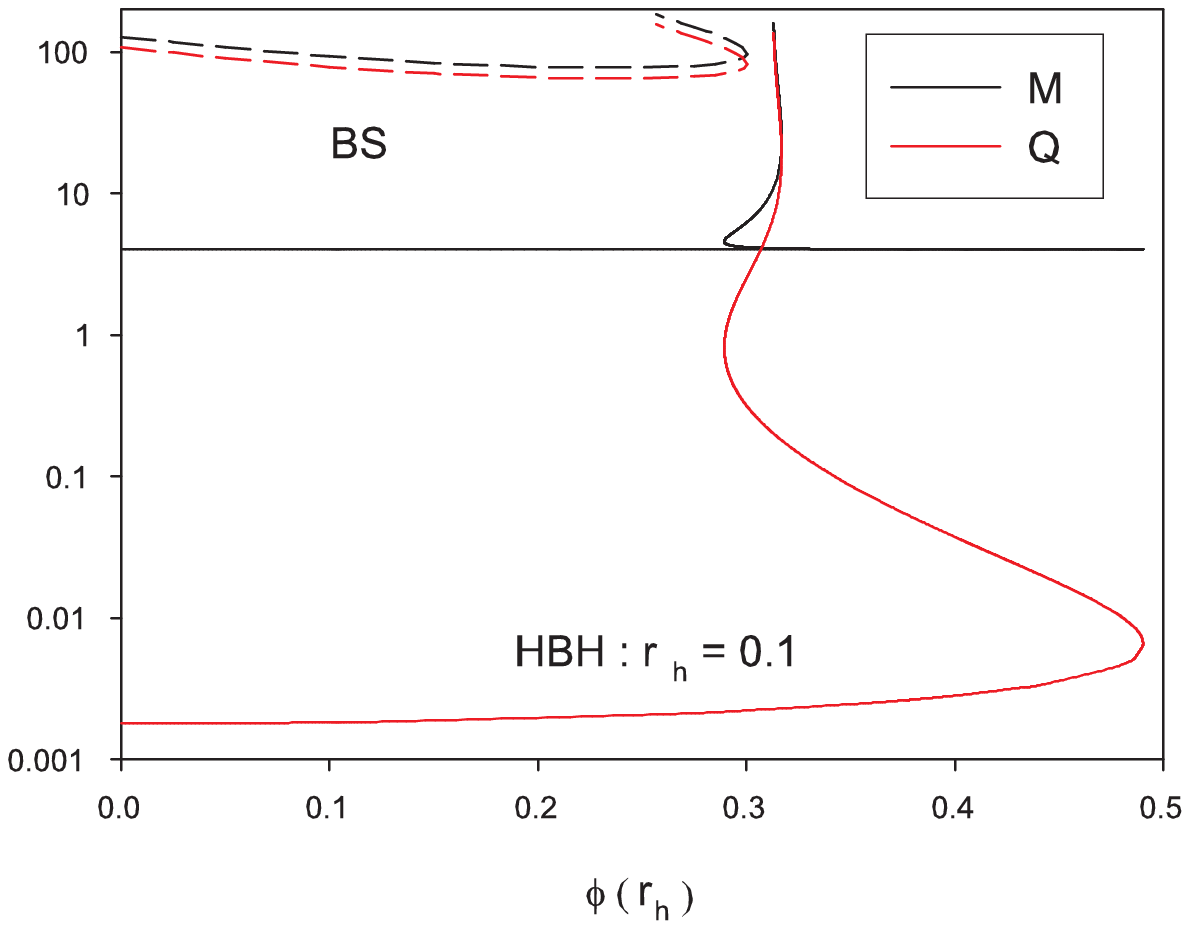}}
{\includegraphics[width=8cm]{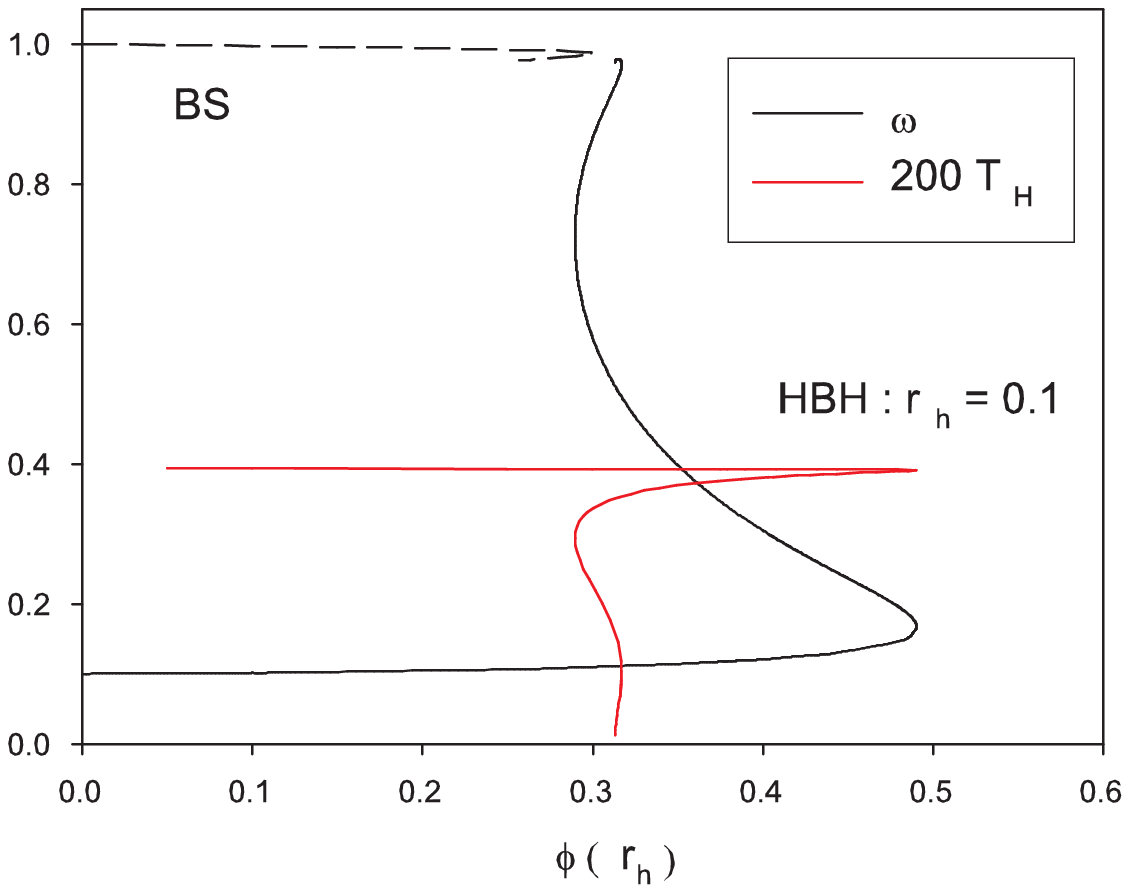}}
\end{center}
\caption{We show the properties of hairy EGB black holes (HBH) as well as of the corresponding boson stars (BS) for 
$\alpha=2.0$ and $q=0.57$. 
Left: The mass $M$ (black) and the charge $Q$ (red) of HBH for $r_h=0.1$  as function of
the value of the scalar field on the horizon $\phi(r_h)$ (dashed) as well as for BS as function of the scalar field at the origin $\phi(0)$ (solid). 
Right: Same as left, but for the Hawking temperature $T_{\rm H}$  and frequency $\omega$.
\label{q_057}
}
\end{figure}
\begin{enumerate}
\item {\bf $q < q_{cr}$}~: 

To demonstrate the behaviour of solutions, we set $q=1/(2\sqrt{2})\approx 0.3535 $. 
The numerical construction confirms that a branch of hairy EGB black holes (HBH) with fixed $r_h$ 
emerges from the charged scalar cloud solution when increasing $\phi(r_h)$. These solutions have  $\omega \geq \omega_{\rm cl}$ and exist only up to a maximal 
value of $\omega=\omega_{\rm m}$. At this maximal value of the frequency, 
a second branch of solutions exist with $\omega < \omega_{\rm m}$.  This is shown in
Fig.~\ref{q_035} for $r_h=0.1$ and $r_h = 0.5$, respectively. Here, the mass $M$ and the charge $Q$
of the HBH are given as function of $\phi(r_h)$ (left) and of $\omega$ (right), respectively. Note that this pattern
of the second branch appears only when using either $\phi(r_h)$ or the frequency $\omega$ as variable. 
The second branch ends at a critical value of $\omega = \omega_{\rm c} \lesssim \omega_{\rm m}$ (see also
the right of  Fig.~\ref{q_035b} for the dependence of $\omega$ on $\phi(r_h)$). 
This limiting configuration has finite mass $M$ and charge $Q$, but the profiles of the solutions present
two regions separated by an intermediate radius $\tilde{r}_h$ with $r_h < \tilde{r}_h < \infty$ for 
which $f(\tilde{r}_h)\to 0$ and $f'(\tilde{r}_h)\to 0$ suggesting that the solution is an extremal black hole solution,
which is confirmed by the fact that the Hawking temperature $T_{\rm H}$ tends to zero (see right of Fig.~\ref{q_035b}). For $r > \tilde{r}_h$ the scalar field is equal to zero and the solution corresponds to the EGB black hole (\ref{egbbh}). However, in the interval $r\in [r_h:\tilde{r}_h]$ the scalar field is non-trivial, while the
potential $V(r)$ is constant. Hence, the now uncharged scalar field becomes trapped between two horizons in a shell with radial thickness $d_s=\tilde{r}_h - r_h$.  
Let us remark that the profile on the left
  side of Fig. \ref{q_035b} corresponds to a HBH with $\alpha= 2.0$, $r_h = 0.5$. In this case, we find 
 $\omega_{\rm c} \approx 0.614$ and $d_s=\tilde{r}_h - r_h \approx 2.138 - 0.5 \approx 1.638$. 

\begin{figure}[ht]
\begin{center}
{\includegraphics[width=8cm]{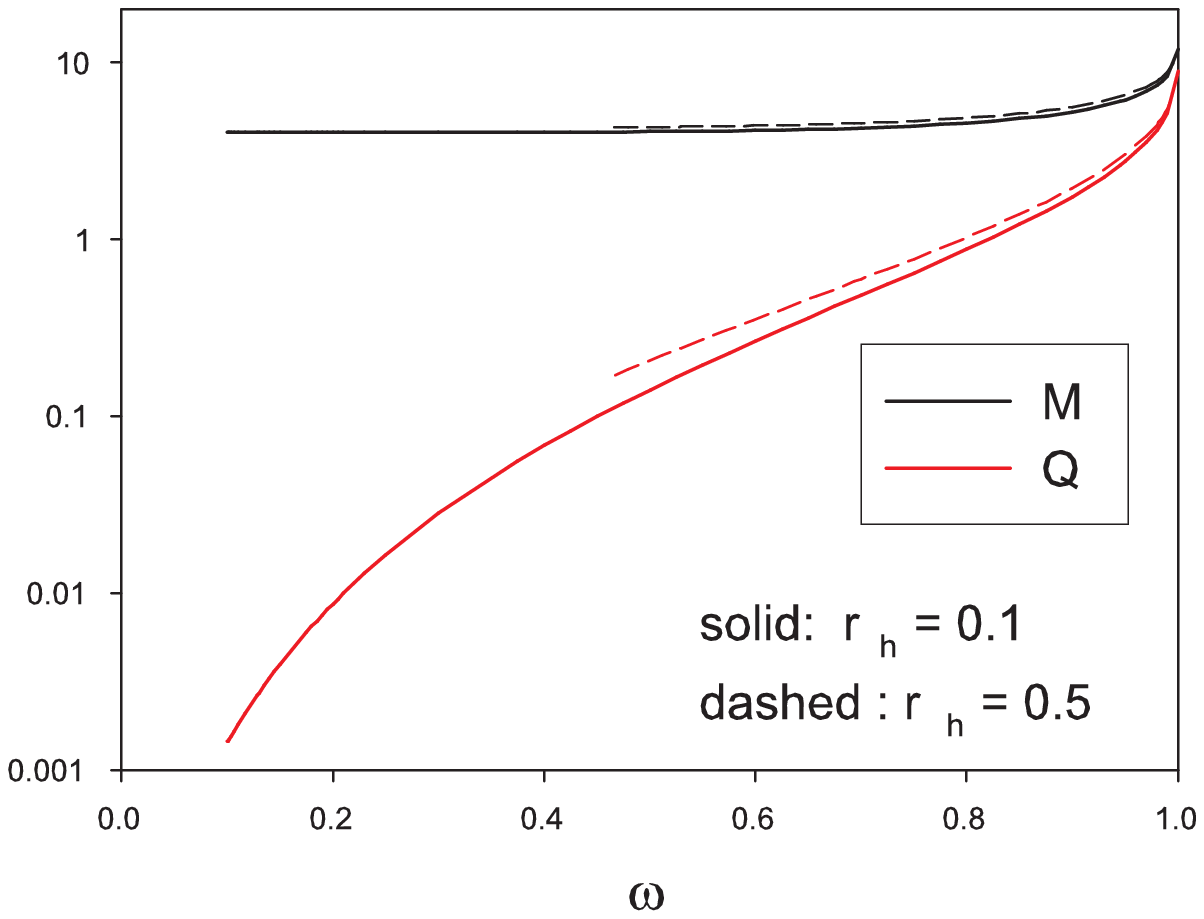}}
{\includegraphics[width=8cm]{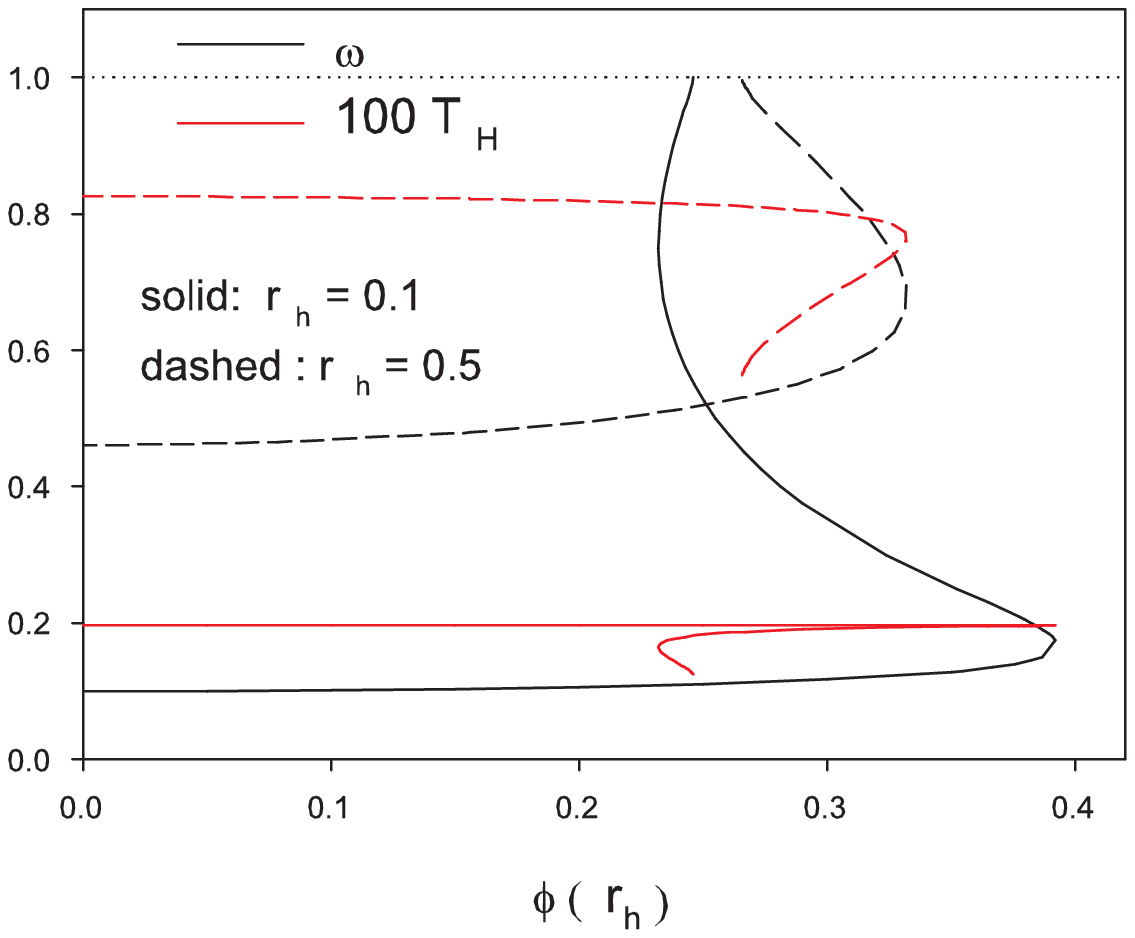}}
\end{center}
\caption{We show the properties of hairy EGB black holes (HBH) as well as of the corresponding boson stars (BS) for 
$\alpha=2.0$ and $q=0.7$. 
Left: The mass $M$ (black) and the charge $Q$ (red) of HBH for $r_h=0.1$ (solid) and $r_h=0.5$ (dashed)  as function of the frequency $\omega$ of the scalar field. 
Right: We show the frequency $\omega$ (black) and the Hawking temperature $T_{\rm H}$  (red) as function of the scalar field
value at the horizon $\phi(r_h)$ for the same parameters. 
\label{q_07}
}
\end{figure}

\item{\bf $q \sim q_{cr}$}~:

As expected from the argument of continuity, the spectrum of HBH evolves smoothly when $q$ is increased.
We find that the hairy black holes exist on a larger interval of the frequency $\omega$ in this case. In particular,
our numerical results suggest that the critical value $\omega_{\rm c}\approx  1$
for $q \to q_{cr}$.
The limiting configuration (similar to the one discussed above for smaller values of $q$) is reached  
after two (or several) backbendings in the parameter $\phi(r_h)$.
This is shown in Fig.~\ref{q_057} for  $q = 0.57$ and $r_h=0.1$, where we give the mass $M$ and the charge $Q$ (left side of
Fig.~\ref{q_057}) as well as the frequency $\omega$ and the Hawking temperature $T_{\rm H}$ (right side of 
Fig.~\ref{q_057}) as function of $\phi(r_h)$. For this particular case, we find three branches of solutions. 

\item{\bf $q > q_{cr}$}~:

Finally, for  $q > q_{cr}$, the numerical results show that the HBH 
exist for $\omega \in [\omega_{{\rm cl}}, 1]$ and that only one branch of solutions in $\omega$ exists.
This is shown in Fig. \ref{q_07}  for  $q = 0.7$ and two different values of the horizon radius, $r_h=0.1$ and $r_h=0.5$, respectively. Here, we
give the mass $M$ and the charge $Q$ as function of the frequency $\omega$ (left side of
Fig.~\ref{q_07}) as well as the frequency $\omega$ and the Hawking temperature $T_{\rm H}$  as function of $\phi(r_h)$ (right side of 
Fig.~\ref{q_07}). For this particular case, we find three branches of solutions in $\phi(r_h)$.
The  limiting behaviour, however, is very different to that in the case $q < q_{cr}$. As is apparent
from Fig.~\ref{q_07} the solution reaches $\omega = 1$ before an extremal HBH is approached. 
Hence, the scalar field ceases to be exponentially localized.  

\end{enumerate}

\section{Conclusions}
In this paper, we have investigated the critical behaviour of the asymptotically flat, charged and non-rotating solutions
available in a 5-dimensional Einstein-Gauss-Bonnet (EGB) gravity model minimally coupled to a $U(1)$ charged scalar field
with harmonic time-dependence. We find that although charged boson stars and hairy EGB black holes
do exist, boson stars do not appear in the limit of vanishing horizon radius for the hairy black holes, but rather co-exist 
with the black holes for the same parameter values. 

Charged scalar field clouds on EGB black holes are limited by the fact that the charged black holes exist up to a maximal
possible charge $Q$, at which they become extremal. The other limit is on the frequency $\omega$ itself, which has to
be smaller than the mass of the scalar field  (for the non self-interacting case) in order to remain exponentially localized.
Using this limit on $\omega$ together with the fact that $\omega$ has to be equal to the upper bound on the superradiant frequency
for scalar clouds to form bound states, the critical limit on $q^2_{\rm cr} = m^2/3$ appears naturally for the extremal non-hairy black hole solution. 
As for other black hole solutions that possess the ability to amplify impigning scalar waves, the hairy black holes appear exactly for $\omega$ larger
than the threshold frequency for the scalar clouds. 

Depending on the choice of $q$ we find that several branches of hairy EGB black holes can exist. These are either limited by the 
condition $\omega^2 \leq m^2\equiv 1$ (for values of $q > q_{\rm cr}$) or by an extremal configuration that possesses 
a shell of uncharged scalar field between the regular (inner) horizon $r_h$ and the extremal (outer) horizon $\tilde{r}_h$ (for $q \leq q_{\rm cr}$), that is, however, an infinite proper distance away from the outside of the extremal black hole.
We plan to investigate the causal space-time structure of these solutions further in the future as they might have interesting implications.


\vspace{2cm} 

{\bf Acknowledgements}
BH would like to thank FAPESP for financial support under grant number {\it 2016/12605-2} and CNPq for financial support under {\it Bolsa de Produtividade Grant 304100/2015-3.}

\clearpage


\clearpage

\section{Appendix~: A no-hair theorem for charged scalar fields with harmonic time-dependence on $d$-dimensional Tangherlini-Schwarzschild and
Reissner-Nordstr\"om solutions}
Here we would like to extend the arguments for the non-existence of scalar hair with harmonic time-dependence on 4-dimensional static, spherically symmetric, asymptotically flat black holes \cite{Pena:1997cy}
to the charged case as well as to higher dimensions.

We use the following spherically symmetric and static metric Ansatz in $d$ dimensions :
\begin{equation}
ds^2 = -b(r) dt^2 + \frac{1}{f(r)} dr^2 + r^2 d\Omega_{d-2}^2 
\end{equation}
with $d\Omega_{d-2}^2 = d\theta_1^2 + \sin^2 \theta_1 d\theta_2^2 + ... 
+ {\displaystyle \prod_{k=1}^{d-3}} \sin^2\theta_k d\theta_{d-2}^2$ the line element of the $(d-2)$-dimensional unit sphere parametrized by the angles $\theta_1$, $\theta_2 ,...,\theta_{d-2}$. 
The components of the Einstein equation $G_{\mu\nu}=8\pi G_{d} T_{\mu\nu}$ then read (see \cite{tangherlini}
for the equations, albeit in a slightly different parametrization of the metric)~:
\begin{equation}
\label{einsteintt}
(d-2) \frac{f}{2} \left(\frac{f'}{rf} + \frac{d-3}{r^2}\right) - \frac{(d-2)(d-3)}{2r^2} = 8\pi G_d T^t_t  \ ,
\end{equation}
\begin{equation}
\label{einsteinrr}
(d-2)\frac{f}{2} \left(\frac{b'}{rb}+ \frac{d-3}{r^2}\right) - \frac{(d-2)(d-3)}{2r^2} = 8\pi G_d T^r_r \   , 
\end{equation}
\begin{equation}
\label{einsteinthetatheta}
\frac{f}{2} \left[\frac{b''}{b} - \frac{b'^2}{2 b^2} + \frac{f'b'}{fb} + \frac{(d-4)(d-3)}{r^2} + \frac{(d-3)}{r}\left(\frac{f'}{f} + \frac{b'}{b}\right)\right] - \frac{(d-3)(d-4)}{2r^2} = 8\pi G_d T^{\theta}_{\theta}
\end{equation}
where $\theta$ in the last component stands for $\theta_1, \theta_2,...\theta_{d-2}$. 
Furthermore, there exists a constraint, which reads~:
\begin{equation}
\label{constraint}
(T^r_r)' = \frac{b'}{2b}\left(T^t_t - T^r_r\right) + \frac{d-2}{r}\left(T^{\theta}_{\theta}-T^r_r\right) \ .
\end{equation}
The {\it weak energy condition} requires that all spatial components $T^i_i$ with $i=1,...,d$ 
fulfil $T^i_i \geq T^t_t$ with $-T^t_t$ equal to the energy density $\rho$.

It is clear from (\ref{constraint}) that we need to require $(T^t_t(r_h)-T^r_r(r_h))/b(r_h)$ to be regular on the horizon $r_h$ for any type of energy-momentum tensor, if we want the $T^r_r$ component
to behave regularly on the horizon. Since $T^t_t = - \rho \leq  0$ for all $r$ and hence in particular at $r=r_h$, this implies automatically $T^r_r(r_h) \leq 0$. 
Moreover, from the requirement of asymptotic flatness we know that $T^r_r (r\rightarrow \infty) \rightarrow 0$. 
Note that using (\ref{einsteintt}) and (\ref{einsteinrr}) this also implies that $f'(r_h)/f(r_h)=b'(r_h)/b(r_h)$. 
Now let us discuss the following two cases~:
\begin{itemize}
\item Electric field with $T^M_M=k\cdot{\rm diag}(-1,-1,+1,+1,...,+1) f V'^2/b$, $M=0,...d-1$, where the constant $k$ depends on the units chosen, but is positive. In this case $T^t_t=T^r_r$ and $T^{\theta}_{\theta}=-T^r_r$ such that (\ref{constraint}) tells
us that the derivative of $T^r_r$ is positive and hence solutions exist. These are the Tangherlini-Reissner-Nordstr\"om solutions \cite{tangherlini}.
\item Fields that fulfil $T^{\theta}_{\theta} \leq T^r_r$. This has been discussed in \cite{Pena:1997cy}
for the 4-dimensional vacuum case and an uncharged scalar field. Using the energy momentum tensor of the charged scalar field given by
\begin{equation}
T_{MN} = - g_{MN} \left((D^{K}\Phi)^* D_K \Phi + V(\Phi)\right) + (D_M \Phi)^* D_N \Phi + D_M \Phi (D_N \Phi)^*    
\ \ , \ \  M,N,K=0,...,d-1
\end{equation}
we find that
\begin{equation}
T^t_t=-\varepsilon_1 - \varepsilon_2 - V \ \ , \ \ T^r_r=\varepsilon_1 + \varepsilon_2 - v \ \ , \ \ T^{\theta}_{\theta}=\varepsilon_1 - \varepsilon_2 - v
\end{equation}
where
\begin{equation}
\varepsilon_1=\frac{1}{b} \left(\omega+qV\right)^2 \phi^2 \ \ , \ \   \varepsilon_2=f \phi'^2 \ \ , \ \ 
V= m^2 \phi^2   \ .
\end{equation}
Hence, the first term on the rhs of (\ref{constraint}) is regular and negative on the horizon remembering that $f'(r_h)/f(r_h)=b'(r_h)/b(r_h)$ and 
$f'(r_h)/(4\pi) = T_{\rm H} \geq  0$, while $T^{\theta}_{\theta} - T^r_r = - 2\varepsilon_2 \leq 0$. 
Hence, the charged scalar field in $d$ dimensions fulfils the criteria and we conclude that {\it charged scalar hair with harmonic
time dependence cannot exist on Schwarzschild or Reissner-Nordstr\"om black holes in $d$ dimensions.}
 
\end{itemize}

 \end{document}